\documentclass[12pt]{spieman}  		
\usepackage{amsmath,amsfonts,amssymb}
\usepackage{graphicx}
\usepackage{setspace}
\usepackage{tocloft}

\def\wisk#1{\ifmmode{#1}\else{$#1$}\fi}
\def\deg    {\wisk{^\circ}}
\def\amin   {\wisk{^\prime\ }}

\title{Systematic error cancellation for a four-port interferometric polarimeter}

\author[a,*]{A. Kogut}
\author[a,b]{D. J. Fixsen}
\affil[a]{Goddard Space Flight Center, Code 665, 
	8800 Greenbelt Road,
	Greenbelt, MD, USA 20771}
\affil[b]{University of Maryland, College Park, MD, USA 20742}

\cftpagenumbersoff{figure}
\cftpagenumbersoff{table} 
\begin{document} 
\maketitle


\begin{abstract}
The Primordial Inflation Explorer (PIXIE) 
is an Explorer-class mission concept 
to measure the gravitational-wave signature of primordial inflation 
through its distinctive imprint on the linear polarization 
of the cosmic microwave background (CMB).  
Its optical system couples a polarizing Fourier transform spectrometer 
to the sky to measure the differential signal 
between orthogonal linear polarization states 
from two co-pointed beams on the sky.  
The double differential nature of the four-port measurement 
mitigates beam-related systematic errors 
common to the two-port systems used in most CMB measurements.  
Systematic errors coupling unpolarized temperature gradients
to a false polarized signal
cancel to first order for any individual detector.
This common-mode cancellation is performed optically,
prior to detection,
and does not depend on the instrument calibration.
Systematic errors coupling temperature to polarization
cancel to second order
when comparing signals from independent detectors.
We describe the polarized beam patterns for PIXIE 
and assess the systematic error 
for measurements of CMB polarization.
\end{abstract}

\keywords{
cosmic microwave background,
systematic error,
beam patterns,
polarimeter,
Fourier transform spectrometer}

{\noindent \footnotesize\textbf{*}
Address all correspondence to:
Alan Kogut,  \linkable{alan.j.kogut@nasa.gov} }


\begin{spacing}{1}   		

\section{Introduction}
Polarization of the cosmic microwave background (CMB)
provides a powerful test of the physics of the early universe.
An arbitrary pattern of linear polarization
mapped over the sky
may be decomposed into
a spatially symmetric component
(even parity E-modes)
and an
anti-symmetric component
(odd parity B-modes).
Scalar sources such as
temperature or density perturbations
can only generate even-parity E-modes,
while
gravitational waves
created during an inflationary epoch
in the early universe
can generate either parity.
Detection of the B-mode signal 
in the CMB polarization field
is thus recognized as a ``smoking gun'' signature of inflation,
testing physics at energies inaccessible through any other means
\cite{
rubakov/etal:1982,
fabbri/pollock:1983,
abbott/wise:1984,
polnarev:1985,
davis/etal:1992,
grishchuk:1993,
kamionkowski/etal:1997,
seljak/zaldarriaga:1997}.

%
\begin{figure}[b]
\centerline{
\includegraphics[height=2.5in]{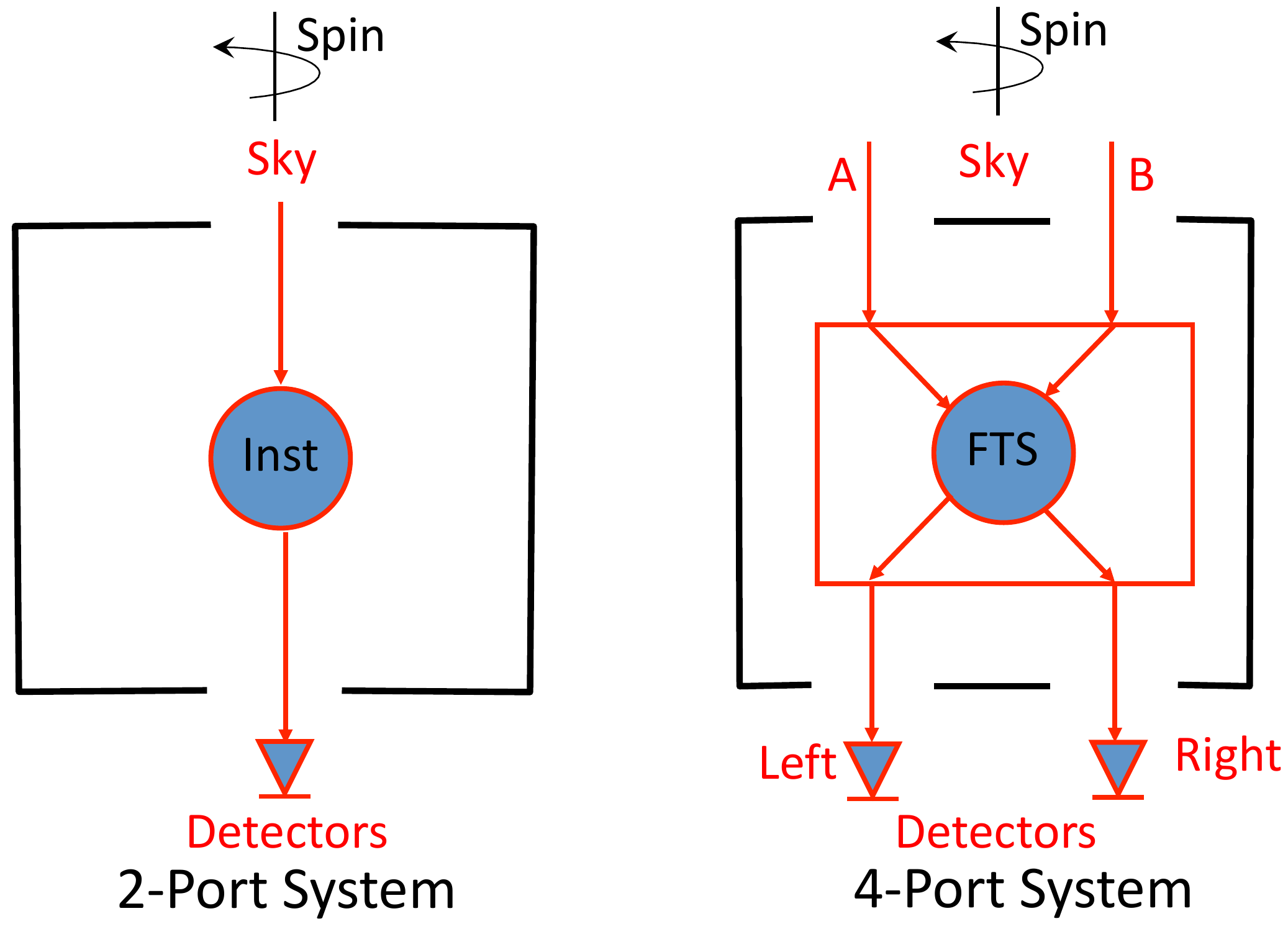}}
\caption[Four-port vs two-port operation]
{A two-port system (left) couples a single linear polarization
from the sky to a single detector.
PIXIE's polarizing Fourier Transform Spectrometer
operates as a four-port device (right)
with two input ports open to the sky
and two output ports terminated by
polarization-sensitive detectors.
Interfering the two beams
cancels the effects of 
common mode beam ellipticity,
as each detector then couples 
to both linear polarizations from the sky.
\label{4_port_fig}}
\end{figure}
%

The amplitude of the gravitational wave signal 
depends on the energy scale of inflation as
\vspace{-2mm}
\begin{equation}
E = 1.06 \times 10^{16} 
\left( 
\frac{r}{0.01} 
\right)^{1/4}
~{\rm GeV} 
\label{potential_eq}
\vspace{-2mm}
\end{equation}
where 
$r$ is the power ratio of gravitational waves to density fluctuations
\cite{lyth/riotto:1999}.
In most large-field models,
$r$ is predicted to be of order 0.01,
corresponding to polarized amplitude 30 nK
or energy near the Grand Unified Theory scale, $10^{16}$ GeV.
Signals at this amplitude
could be detected by  a dedicated polarimeter,
providing a critical test of a central component of modern cosmology.
Detection of a gravitational-wave component in the CMB polarization
would have profound implications for both cosmology and high-energy physics.
It would provide strong evidence for inflation,
provide a direct, model-independent determination
of the relevant energy scale,
and test physics at energies a trillion times beyond those
accessible to particle accelerators.
Generation of gravitational waves during inflation is purely a
quantum-mechanical process:
a detection of the B-mode signal
provides direct observational evidence
that gravity obeys quantum mechanics.

Characterizing the CMB to measure polarization 
at the parts-per-billion level
requires careful control of systematic errors.
A particular concern
are systematic errors related to the instrument optics,
which can couple
the much brighter unpolarized temperature fluctuations
into a false polarization signal.
All CMB instruments must couple the detectors to the sky,
and must therefore account for potential 
beam-related systematic errors.
An extensive literature discusses
common effects and mitigation strategies
\cite{
hu/etal:2003,
odea/etal:2007,
rosset/etal:2007,
shimon/etal:2008}.

The Primordial Inflation Explorer (PIXIE)
is an Explorer-class mission
designed to measure  
the inflationary signature in polarization
as well as 
distortions from the 2.725 K blackbody spectrum
induced by energy-releasing processes
at more recent cosmological epochs
\cite{kogut/etal:2011}.
Its projected sensitivity of a few nK on degree angular scales
or larger
corresponds to a limit
$r < 10^{-3}$ at 5 standard deviations.
PIXIE differs from most CMB polarimeters
in its use of a polarizing Fourier transform spectrometer
coupled to the sky through a
multi-moded optical system.
The double differential nature of the resulting four-port measurement 
minimizes beam-related systematic errors 
common to the two-port systems used in most CMB measurements.  
We describe the polarized beam patterns for PIXIE 
and assess the systematic error 
for measurements of CMB polarization.

\section{PIXIE Optical System}
A common implementation for CMB polarimetry
images the sky onto a set of polarization-sensitive detectors.
Since each detector is sensitive to a single linear polarization
from the sky
(although two or more detectors may share a physical pixel),
the resulting system may be described as a two-port device
with any polarization comparison between detectors 
occurring post-detection.
In contrast,
the PIXIE optical system forms a four-port device
(Fig \ref{4_port_fig}).
Reflective optics couple a
polarizing Fourier Transform Spectrometer (FTS) to the sky.
The FTS introduces an optical phase delay
between the two input beams,
and routes recombined beams
to non-imaging concentrators
at each of two output ports.
Within each concentrator,
a pair of polarization-sensitive detectors
measure the power as a function of optical phase delay.
Let $\vec{E} = E_x \hat{x} + E_y \hat{y}$ 
represent the electric field incident from the sky.
The power $P$ at the detectors
as a function of the phase delay $z$
may be written
\begin{eqnarray}
P_{Lx} &=& 1/2 ~\smallint \{ ~(E_{Ax}^2+E_{By}^2)+(E_{Ax}^2-E_{By}^2) \cos(4z\omega /c) ~\}d\omega   \nonumber \\
P_{Ly} &=& 1/2 ~\smallint \{ ~(E_{Ay}^2+E_{Bx}^2)+(E_{Ay}^2-E_{Bx}^2) \cos(4z\omega /c) ~\}d\omega   \nonumber \\
P_{Rx} &=& 1/2 ~\smallint \{ ~(E_{Ay}^2+E_{Bx}^2)+(E_{Bx}^2-E_{Ay}^2) \cos(4z\omega /c) ~\}d\omega    \nonumber \\
P_{Ry} &=& 1/2 ~\smallint \{ ~(E_{Ax}^2+E_{By}^2)+(E_{By}^2-E_{Ax}^2) \cos(4z\omega /c) ~\}d\omega~,
\label{full_p_eq}
\end{eqnarray}
where
$\hat{x}$ and $\hat{y}$ refer to orthogonal linear polarizations,
L and R refer to the detectors in the left and right concentrators,
A and B refer to the two input beams,
$\omega$ is the angular frequency of incident radiation,
and the factor of 4 reflects the symmetric folding of the optical path.
When both input ports are open to the sky,
the power at each detector consists of a dc term 
proportional to the intensity 
$E_x^2 + E_y^2$
(Stokes $I$)
plus a term 
modulated by the phase delay $z$,
proportional to the linear polarization 
$E_x^2 - E_y^2$
(Stokes $Q$) 
in instrument-fixed coordinates.
Each detector is thus sensitive to 
the difference between orthogonal linear polarizations
from the two input ports,
with the difference now occurring pre-detection.
Rotation of the instrument about the beam axis
rotates the instrument coordinate system
relative to the sky
to allow separation of Stokes $Q$ and $U$ parameters on the sky.

Rotation of the instrument relative to the sky
can produce systematic errors in the recovered polarization
if the instrument beams are not azimuthally symmetric.
This effect has been well studied
for two-port devices
which couple the sky directly to 
a single polarization-sensitive detector.
The dominant systematic error for such a device
is temperature--polarization coupling
as the beam ellipticity
interacts with local gradients in the unpolarized sky intensity,
producing a spin-dependent signal
degenerate with true polarization.
Temperature--polarization coupling
can be mitigated in hardware
using such techniques as polarization modulation
(rapidly switching a single detetor 
between orthogonal polarization states)
or in analysis using a well-measured beam profile.

\begin{figure}[b]
\begin{center}
\begin{tabular}{c}
\includegraphics[height=3.5in]{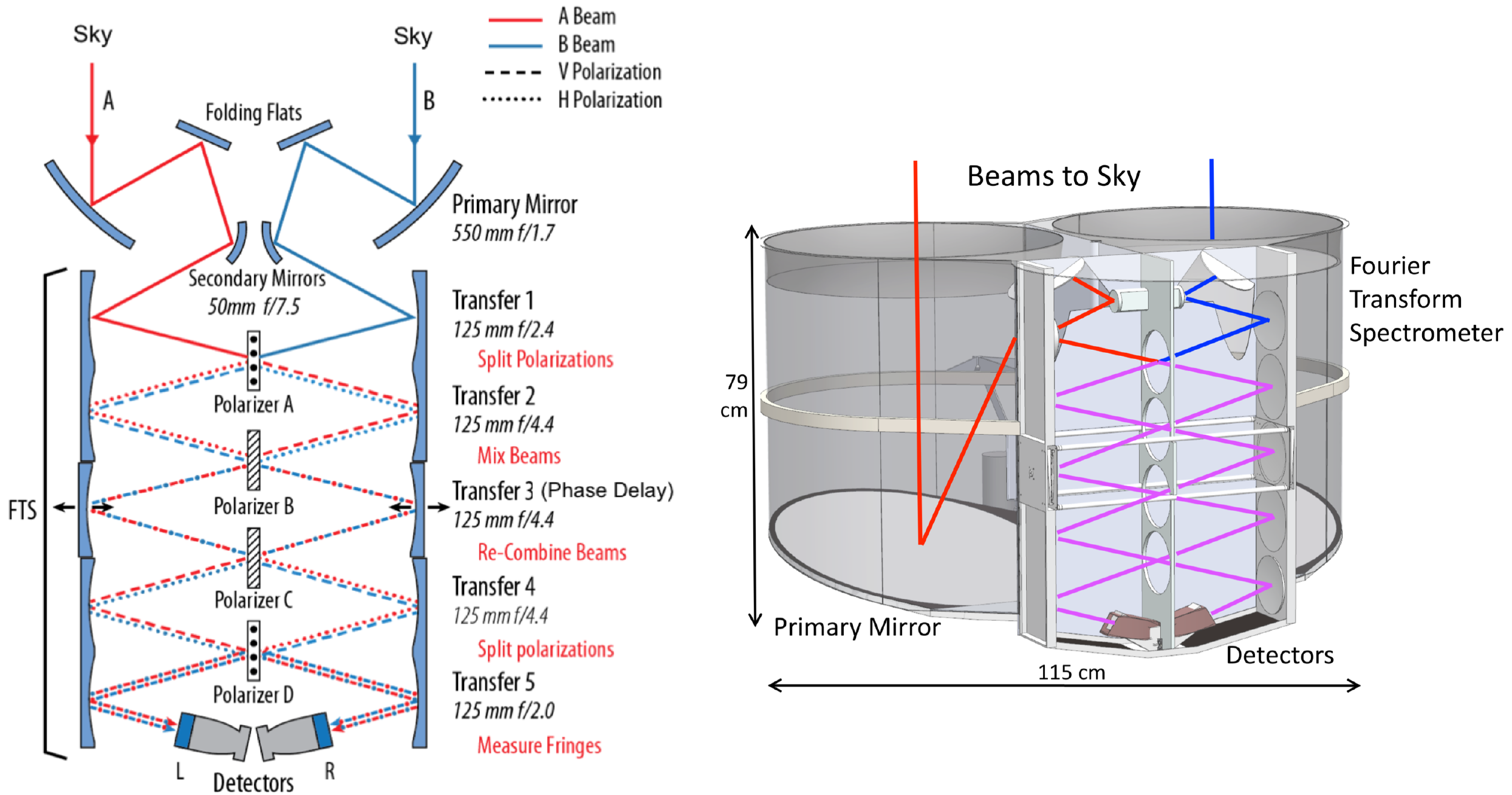}
\end{tabular}
\end{center}
\caption
{PIXIE optical signal path.  
The left panel shows the optical elements within the
Fourier Transform Spectrometer
while the right panel shows the physical layout.
\label{pixie_optics_layout}
}
\end{figure}

The beam response of a four-port system 
is considerably more complicated
than the two-port systems commonly used for CMB measurements.
Although a four-port system may still employ
mitigation strategies such as rotation or beam mapping, 
the double differential nature of the four-port measurement
provides additional mitigation 
against common spin-dependent polarization errors
while simultaneously providing a means to identify and correct
residual effects.
Figure \ref{pixie_optics_layout} shows the PIXIE optical path.
Consider (in a time-reversed sense)
the path through the optics
taken by photons leaving the 
$\hat{x}$ detector in the left-side concentrator.
Since this detector is sensitive to a single linear polarization,
the photons exiting the left-side concentrator
are entirely in the $\hat{x}$ polarization.
A series of polarizing wire grids within the FTS
splits the beam and rotates the polarization
so that half the initial power
exits through port A in the $\hat{y}$ polarization
while the other half
exits through port B in the $\hat{x}$ polarization
(see, e.g., Appendix A of reference \citenum{kogut/etal:2011}).
A set of reflective mirrors
then couples ports A and B to the sky
while preserving the polarization state.
Stops at transfer mirror 5
and at the entrance to the concentrator
circularize the beam.
The entire instrument, including all baffling,
remains isothermal at 2.725 K
so that rays scattered out of the beam
terminate on an isothermal black surface.
Such rays contribute to the photon noise budget
but do not introduce artifacts in the beams.

Let us define the beam pattern of the concentrator as
$H_x(\theta, \phi)$ for the $\hat{x}$ polarization
and
$H_y(\theta, \phi)$ for the $\hat{y}$ polarization,
where the angular coordinates
$\theta$ and $\phi$
are referred to the sky.
Similarly, we define the beam pattern for the fore-optics 
(defined as all elements in the optical chain
skyward of the concentrator feed)
as
$F_x(\theta, \phi)$ 
and 
$F_y(\theta, \phi)$.
Using subscripts 
$L$ and $R$ to distinguish the two concentrator ports
and
$A$ and $B$ for the two fore-optic ports,
we may re-write Equation \ref{full_p_eq} as
\begin{eqnarray}
P_{Lx} &\propto& H_{Lx} \left[ F_{Ax} E_{x}^2 - F_{By} E_{y}^2 \right]   \nonumber \\
P_{Ly} &\propto& H_{Ly} \left[ F_{Ay} E_{y}^2 - F_{Bx} E_{x}^2 \right]   \nonumber \\
P_{Rx} &\propto& H_{Rx} \left[ F_{Bx} E_{x}^2 - F_{Ay} E_{y}^2 \right]   \nonumber \\
P_{Ry} &\propto& H_{Ry} \left[ F_{By} E_{y}^2 - F_{Ax} E_{x}^2 \right] ,
\label{full_beam_eq}
\end{eqnarray}
where for clarity we suppress the
dependence on angular coordinates $(\theta, \phi)$
as well as the phase delay integral over frequency.
Two points are apparent.
First, the signal at any single detector
depends on the 
convolution of the concentrator beam profile with the
{\it differential} beam profile
generated by the A- and B-side fore-optics.
To the extent that the A- and B-side optics have identical beam patterns,
the detectors produce no response from an unpolarized sky,
regardless of the intensity gradient on the sky
or the ellipticity of the fore-optics.
This common mode cancellation is performed optically,
prior to detection,
and does not depend on the instrument calibration.
Second, the beam pattern for the concentrator horn
appears only as a common-mode multiplicative factor.
Systematic errors 
coupling temperature anisotropy to polarization
thus cancel to second order
when comparing signals
from independent detectors.

\begin{figure}[t]
\begin{center}
\includegraphics[width=6.0in]{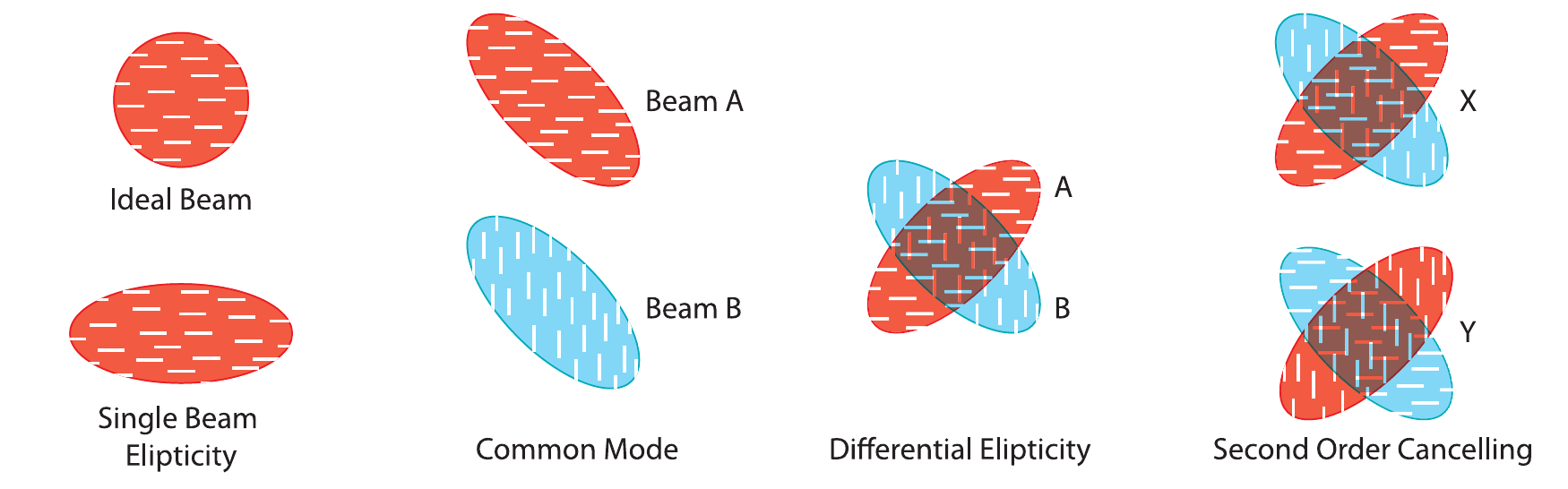}
\end{center}
\caption
{Cartoon illustrating signal cancellation
from differential beam profiles. 
Colored regions indicate the beam shape,
while the white lines indicate the polarization state
accepted by each beam.
\label{common_mode_fig}
}
\end{figure}

Figure \ref{common_mode_fig} illustrates the multiple levels
of common-mode subtraction.
An ideal azimuthally symmetric beam 
would introduce no temperature-polarization coupling.
Real beams, however, will have some ellipticity (left column).
Rotation of an elliptical beam
couples to unpolarized gradients in the sky
to produce a time-dependent signal
degenerate with a true polarization signal.
If, however, 
two beams sensitive to opposite polarization states
but with the same ellipticity are compared,
the common-mode ellipticity cancels 
for unpolarized emission,
leaving no net temperature-polarization coupling
(second column).
Only the \emph{differential} ellipticity
produces a net temperature-polarization coupling,
which appears at second order in the beam difference.
The PIXIE optics employ such beam cancellation
in the $A-B$ comparison for a single detector
(third column, with the differential ellipticity greatly exaggerated).
Comparisons between different detectors
provide an additional level of cancellation.
Each concentrator
contains two detectors sensitive to 
orthogonal polarization states
(Eq. \ref{full_p_eq}),
which view the same sky through the
same fore-optics.
If the differential ellipticity
between the $A$ and $B$ sky beams
is the same for the $\hat{x}$ polarization
as for the $\hat{y}$ polarization,
the net temperature-polarization coupling 
for the single-detector output
will cancel to \emph{second} order in the detector-pair difference
(right-most column).
Alternatively, an orthogonal linear combination of detector pairs
can be chosen
to cancel the sky signal,
thereby isolating any beam effects.
Such measurements can be used
both as confirmation of the expected amplitude of the beam differences 
and to correct residual beam effects in the sky data.

The double-differential beam cancellation
of PIXIE's four-port optical system
reduces the sensitivity to unpolarized gradients on the sky.
The following sections
use Monte Carlo ray-trace code
to evaluate the
common-mode and differential beam patterns.
We quantify the expected systematic error response
for ideal optics
and
show the minimal degradation in performance
after accounting for 
machining and assembly tolerances.

\begin{figure}[t]
\begin{center}
\includegraphics[width=4.0in]{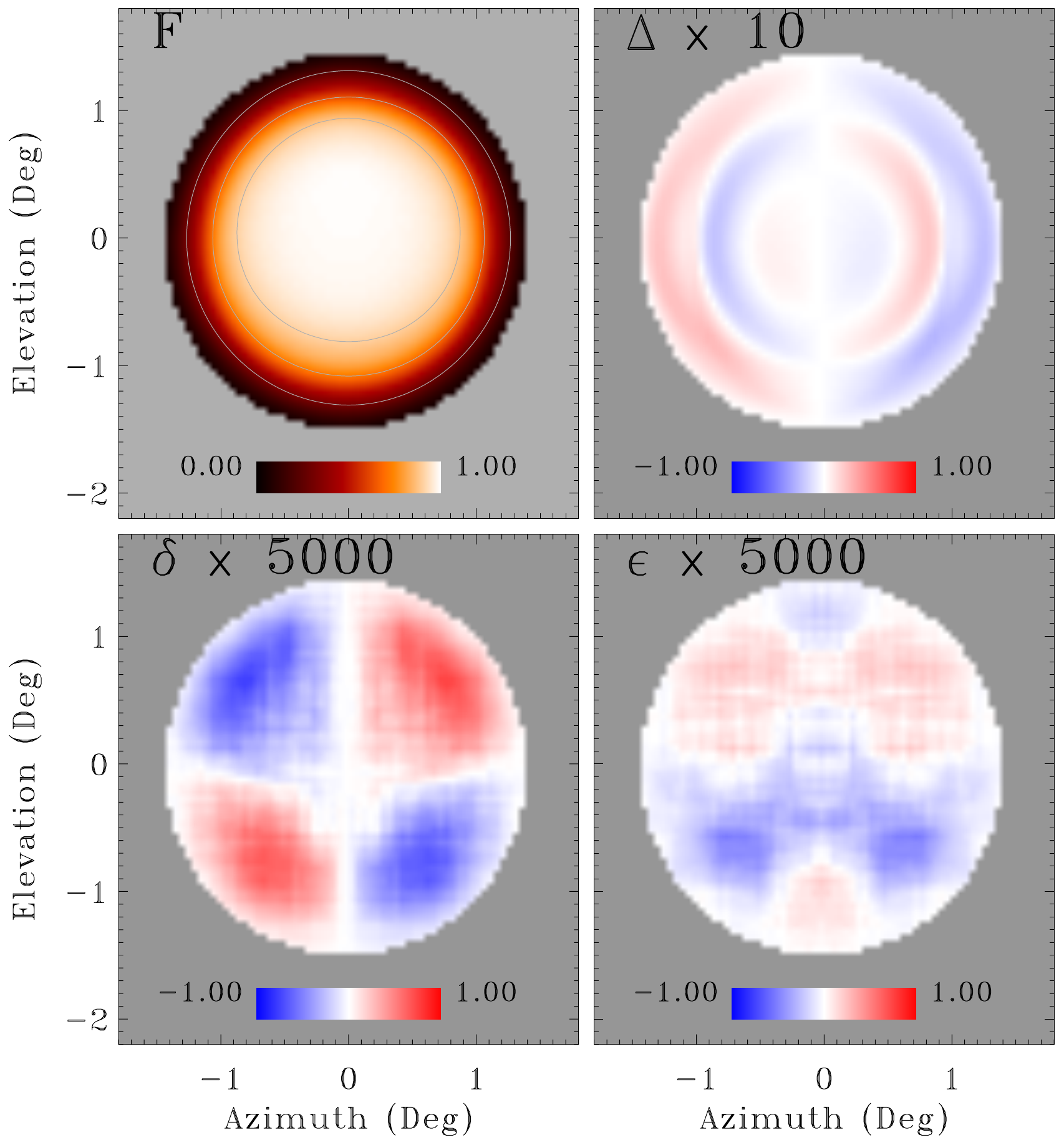}
\end{center}
\caption
{Linear combinations of the PIXIE fore-optics
showing the common-mode and differential beam patterns.
The spatial ($\Delta$)
and polarization ($\delta$)
asymmetries
are small compared to the mean beam pattern $F$.
Contours for the common-mode response $F$
are shown at amplitude 0.3, 0.7, and 0.9
to highlight the circular tophat beam structure.
Note the change in scale for the three differential beam patterns.
\label{fore_optics_beams}
}
\end{figure}

\section{Single-Detector Response}
Systematic errors in the PIXIE four-port optical system 
depend on successive differences in the beam patterns.
We may write the individual fore-optics beam patterns 
in terms of the linear combinations
\begin{eqnarray}
F        &=& \left( F_{Ax} + F_{Ay} + F_{Bx} + F_{By} \right) / 4  	\nonumber \\
\Delta   &=& \left( F_{Ax} + F_{Ay} - F_{Bx} - F_{By} \right) / 4	\nonumber \\
\delta   &=& \left( F_{Ax} - F_{Ay} + F_{Bx} - F_{By} \right) / 4	\nonumber \\
\epsilon &=& \left( F_{Ax} - F_{Ay} - F_{Bx} + F_{By} \right) / 4	
\label{F_def}
\end{eqnarray}
to distinguish the common-mode beam pattern
$F = F(\theta, \phi)$
from the differential beam patterns
$\Delta$ (A--B spatial asymmetry),
$\delta$ ($\hat{x} - \hat{y}$ polarization asymmetry),
and 
$\epsilon$ (spatial/polarization cross term).
With these definitions,
the individual beam patterns become
\begin{eqnarray}
F_{Ax} &=& F + \delta + \Delta + \epsilon \nonumber	\\
F_{Ay} &=& F - \delta + \Delta - \epsilon \nonumber	\\
F_{Bx} &=& F + \delta - \Delta - \epsilon \nonumber	\\
F_{By} &=& F - \delta - \Delta + \epsilon ~.
\label{FAX_def}
\end{eqnarray}
Note that these four linear combinations 
represent a complete set,
carrying all information
for 2 ports in 2 linear polarizations.

Figure \ref{fore_optics_beams}
shows the common-mode and differential beam patterns,
using a Monte Carlo ray-trace code to propagate $10^{11}$ rays
through the PIXIE fore-optics.
As expected, the
beams are dominated by the common-mode illumination $F$.
Since $F$ by definition is the average of the beams,
it can not generate any \emph{differential} ellipticity
and thus can not generate temperature-polarization coupling
regardless of its azimuthal structure.

\begin{figure}[b]
\begin{center}
\includegraphics[width=3.0in]{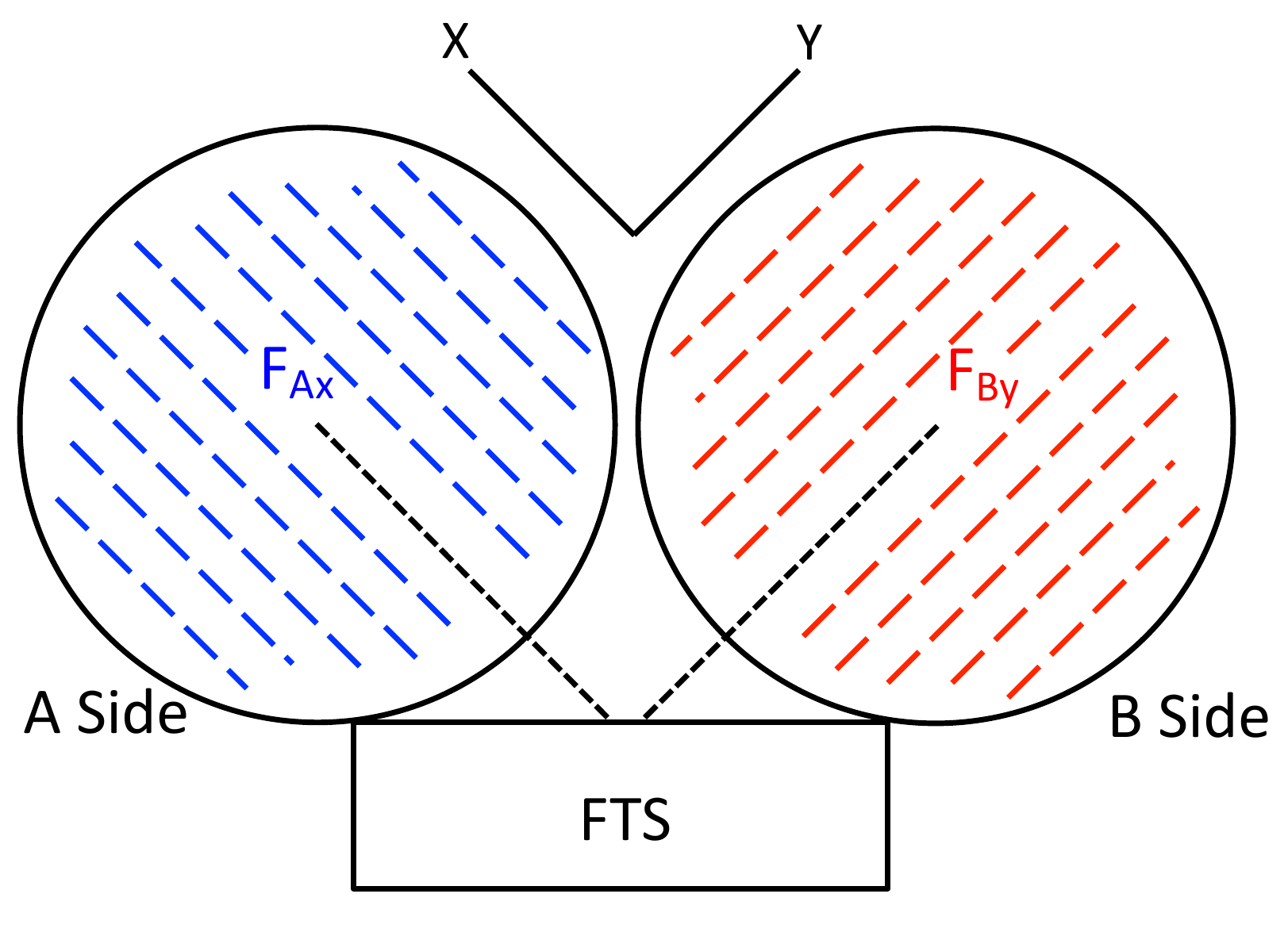}
\end{center}
\caption{ 
Schematic of the PIXIE optical system showing the 
symmetric polarization response
at the beam apertures.
The Fourier transform spectrometer
interferes a single linear polarization
from one side of the instrument
with the orthogonal polarization 
from the other side.
By construction, the
$\hat{x}$ polarization on the A side
is simply the mirror reflection of the
$\hat{y}$ polarization on the B side.
\label{pol_symmetry_cartoon}
}
\end{figure}

Differences between the left and right beams are
are measured by the $A-B$ spatial asymmetry $\Delta(\theta, \phi)$.
Out-of-plane reflections
at the secondary mirror and folding flat
generate a dipolar modulation in $\Delta$
with rms amplitude 0.015 of the common-mode beam pattern.
This is the largest differential mismatch between the beams.
The instrument is symmetric about the left-right midplane
so that,
by design, 
the $A$ and $B$ beams are mirror images of each other
(Figure \ref{pol_symmetry_cartoon}).
Structure within one of the beams
will thus be reflected left-to-right in the other beam,
maximizing the net effect along the left-right direction.

Differences between polarization states
are measured by the polarization asymmetry $\delta(\theta, \phi)$.
Since both the $\hat{x}$ and $\hat{y}$ polarizations
from the detectors
are launched at 45\deg
~relative to the symmetry plane of the instrument
($\S$4),
this term is small
(of order $10^{-4}$ of the common-mode pattern).
For completeness,
there is also a spatial/polarization cross term $\epsilon(\theta, \phi)$.
This term is also small (of order $10^{-4}$).
As shown below,
it does not couple to temperature-polarization mixing,
but appears as a small perturbation on the amplitude
of the measured polarization signal.

Using these definitions, it is straightforward 
(if somewhat tedious)
to show that
\begin{eqnarray}
P_{Lx} &=& H_{Lx} [\, \, ~ ~~~{\boldsymbol Q} \, F
	~ + ~{\boldsymbol Q} \epsilon			
	~ + ~{\boldsymbol I} ( \, \, ~~\delta + \Delta )	~~]	\nonumber \\
P_{Ly} &=& H_{Ly} [~ -{\boldsymbol Q} \, F			
	~ + ~{\boldsymbol Q} \epsilon			
	~ + ~{\boldsymbol I} (  -\delta + \Delta )	~~]	\nonumber \\	
P_{Rx} &=& H_{Rx} [\, \, ~ ~~~{\boldsymbol Q} \, F				
	~ - ~{\boldsymbol Q} \epsilon			
	~ + ~{\boldsymbol I} (\, \,  ~~\delta - \Delta )	~~]	\nonumber \\	
P_{Ry} &=& H_{Ry} [~ -{\boldsymbol Q} \, F			
	~ - ~{\boldsymbol Q} \epsilon	
	~ + ~{\boldsymbol I} ( -\delta - \Delta )	~~]	~.
\label{4_det_with_delta}
\end{eqnarray}

\noindent
The first term in brackets represents the desired polarized sky signal
${\boldsymbol Q(\theta, \phi)}$,
convolved with the mean fore-optics beam pattern.
The second term, 
${\boldsymbol Q} \, \epsilon$,
convolves the true sky polarization 
with the cross beam pattern $\epsilon(\theta, \phi)$.
This term is small.
The cross beam pattern may be written
as the double difference
\begin{equation}
\epsilon = (F_{Ax} - F_{Ay}) - (F_{Bx} - F_{By})
\label{eps_def}
\end{equation}
and is thus second order in the beam difference.
Furthermore, since this term does not mix 
the Stokes parameter ${\boldsymbol Q}$
with either ${\boldsymbol U}$ or ${\boldsymbol I}$,
it only appears as a scale error 
in the amplitude of the true sky polarization
and may be absorbed by the calibration.
The final term
represents systematic temperature--polarization coupling.

The left--right symmetry of the PIXIE optics minimizes 
temperature-polarization coupling.
PIXIE's optical design interferes the $\hat{x}$ polarization 
from one beam
with the $\hat{y}$ polarization from the other beam
(Eq. \ref{full_beam_eq}).
The optical system is symmetric about the central plane,
so that
the $\hat{x}$ polarization from one beam
is the mirror reflection
of the 
$\hat{y}$ polarization from the other beam
(Fig \ref{pol_symmetry_cartoon}).
This enforces a reflection symmetry
such that
\begin{eqnarray}
F_{Ax}(\theta, \phi) &=& F_{By}(\theta, -\phi)	\nonumber \\
F_{Ay}(\theta, \phi) &=& F_{Bx}(\theta, -\phi)	
\label{F_symmetry_eq}
\end{eqnarray}
where the azimuthal angle $\phi$
is defined from the midline.
Note that this left--right symmetry
is not equivalent to an $\hat{x}$--$\hat{y}$ symmetry
since the $\hat{x}$--$\hat{y}$ coordinate system
is rotated by 45\deg~with respect to the optical midline.
Temperature-polarization mixing thus depends on
the linear combinations
\begin{eqnarray}
\delta + \Delta &=& F_{Ax} - F_{By}	\nonumber \\
\delta - \Delta &=& F_{Bx} - F_{Ay}
\label{delta_diff}
\end{eqnarray}
proportional to the
anti-symmetric component of the {\it difference}
between the beams.

The spacecraft spin combines with the 
mirror symmetry of the instrument optics
to further minimize temperature-polarization coupling.
Each detector is sensitive to a single linear polarization
(Stokes $Q$ in a coordinate system
fixed with respect to the instrument).
The entire spacecraft rotates about the instrument boresight
to interchange the roles of $\hat{x}$ and $\hat{y}$ polarization
at the detectors,
allowing full characterization
of the Stokes $Q$ and $U$ parameters on the sky.
True sky polarization is modulated
at twice the spacecraft spin frequency,
\begin{equation}
Q_{\rm inst} = Q_{\rm sky} \cos(2 \gamma) + U_{\rm sky} \sin(2 \gamma) 
\label{qu_sky}
\end{equation}
where $\gamma$ is the spin angle of the instrument
with respect to the sky.
Temperature--polarization mixing is dominated by
the anti-symmetric component of the differential beam pattern
from the instrument fore-optics.
Anti-symmetric signals can only appear at
odd harmonics of the spacecraft spin,
and may readily be distinguished from
true sky polarization.

We quantify the suppression
of temperature--polarization systematic errors
using the spin-dependent moments of the differential beam patterns.
The instantaneous power at each detector
depends on the convolution of the beam pattern
(in instrument-fixed coordinates)
with the sky signal
(rotated from sky to instrument coordinates).
Azimuthal asymmetry in the beam patterns
causes the measured power to vary 
with the spacecraft spin angle.
We thus compute the coefficients
\begin{eqnarray}
a_m &=& \int B(\Omega) \cos(m \phi) d\Omega	\nonumber	\\
b_m &=& \int B(\Omega) \sin(m \phi) d\Omega	
\label{am_def}
\end{eqnarray}
where $B$ represents one of the linear combinations of beam patterns
(Eq. \ref{F_def})
and
$d \Omega = \sin(\theta) d\theta d\phi$
is computed in instrument coordinates
centered on the boresight.

\begin{figure}[t]
\begin{center}
\begin{tabular}{c}
\includegraphics[height=2.6in]{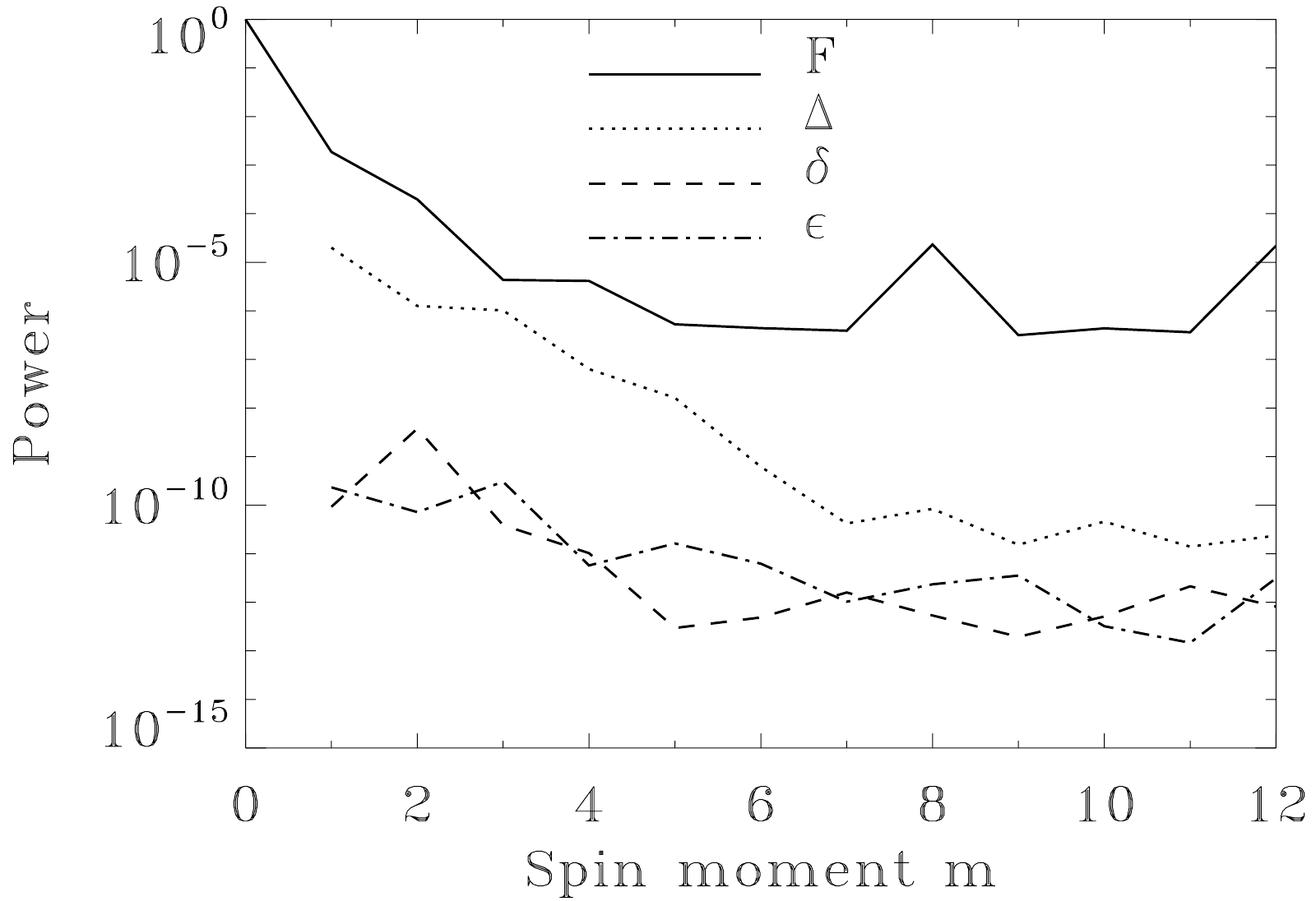}
\end{tabular}
\end{center}
\caption
{Decomposition of the PIXIE differential beam patterns by spin angle.
The common-mode beam $F$ is sensitive only to polarized emission
and does not contribute to temperature--polarization systematic errors.
The mirror symmetry of the PIXIE optics
suppresses temperature--polarization mixing from the 
A--B spatial asymmetry (beam $\Delta$)
by a factor of $10^{-6}$
(see text).
\label{fore_optics_vs_m}
}
\end{figure}

Figure \ref{fore_optics_vs_m}
shows the power $P_m = a_m^2 + b_m^2$
as a function of spin moment $m$.
The odd-even asymmetry in spin moment $m$
is superposed atop an overall decrease in power with increasing $m$.
The noise floor at $P \approx 10^{-12}$
reflects shot noise from the
discrete ray-trace simulation.
Recall that the common-mode beam pattern $F$
is sensitive only to polarized emission on the sky
(Eq. \ref{4_det_with_delta})
and does not create temperature--polarization errors
even for the $m=2$ case.
Systematic errors from temperature--polarization coupling
are dominated by the 
$m=2$ mode of the A--B spatial asymmetry $\Delta$,
and are suppressed by a factor $10^{-6}$
relative to the polarization response in the common-mode beam.

\section{Additional Symmetries}

The mirror symmetry of PIXIE's differential 4-port interferometer
suppresses systematic errors
from temperature--polarization coupling
by 6 orders of magnitude
for the single-detector response.
Additional symmetries between different detectors
allow further suppression 
and identification
of beam-related systematic errors.
The left and right concentrators
are identical,
resulting in left--right symmetry
\begin{eqnarray}
H_{Lx}(\theta, \phi) &=& H_{Rx}(\theta, -\phi)	\nonumber \\
H_{Ly}(\theta, \phi) &=& H_{Ry}(\theta, -\phi)	
\label{H_symmetry_eq}
\end{eqnarray}
for identical polarization states.
This is similar to the left--right symmetry in Eq. \ref{F_symmetry_eq}
except that the symmetry is now between 
identical polarization states on opposite sides of the instrument.

\begin{figure}[b]
\begin{center}
\includegraphics[width=2.5in]{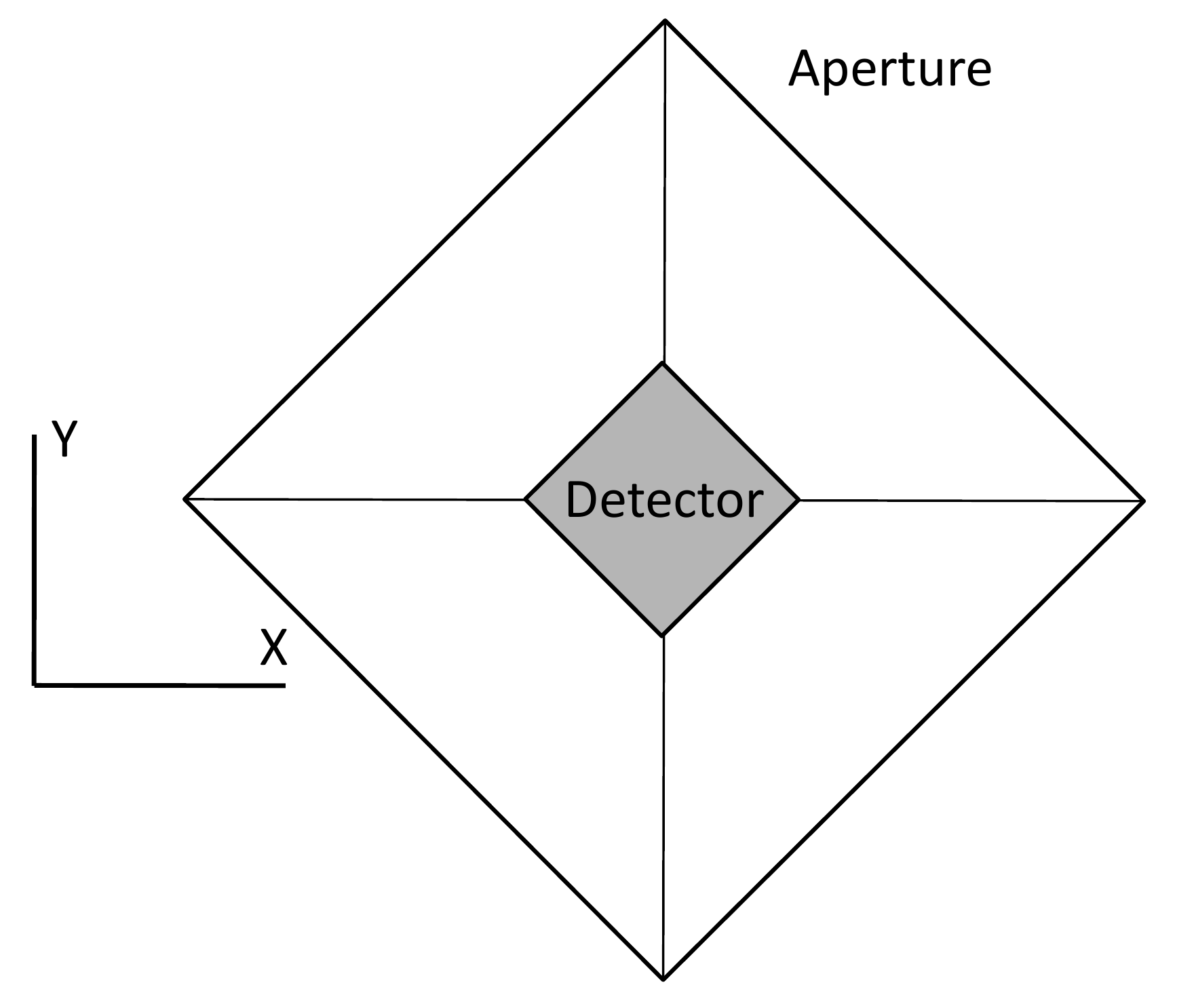}
\end{center}
\caption
{\small 
Schematic showing the orientation of the PIXIE concentrator.
The square aperture is rotated 45\deg
to minimize any differences between the
$\hat{x}$ and $\hat{y}$ polarization.
\label{horn_cartoon}
}
\end{figure}

\begin{figure}[t]
\begin{center}
\includegraphics[width=4.0in]{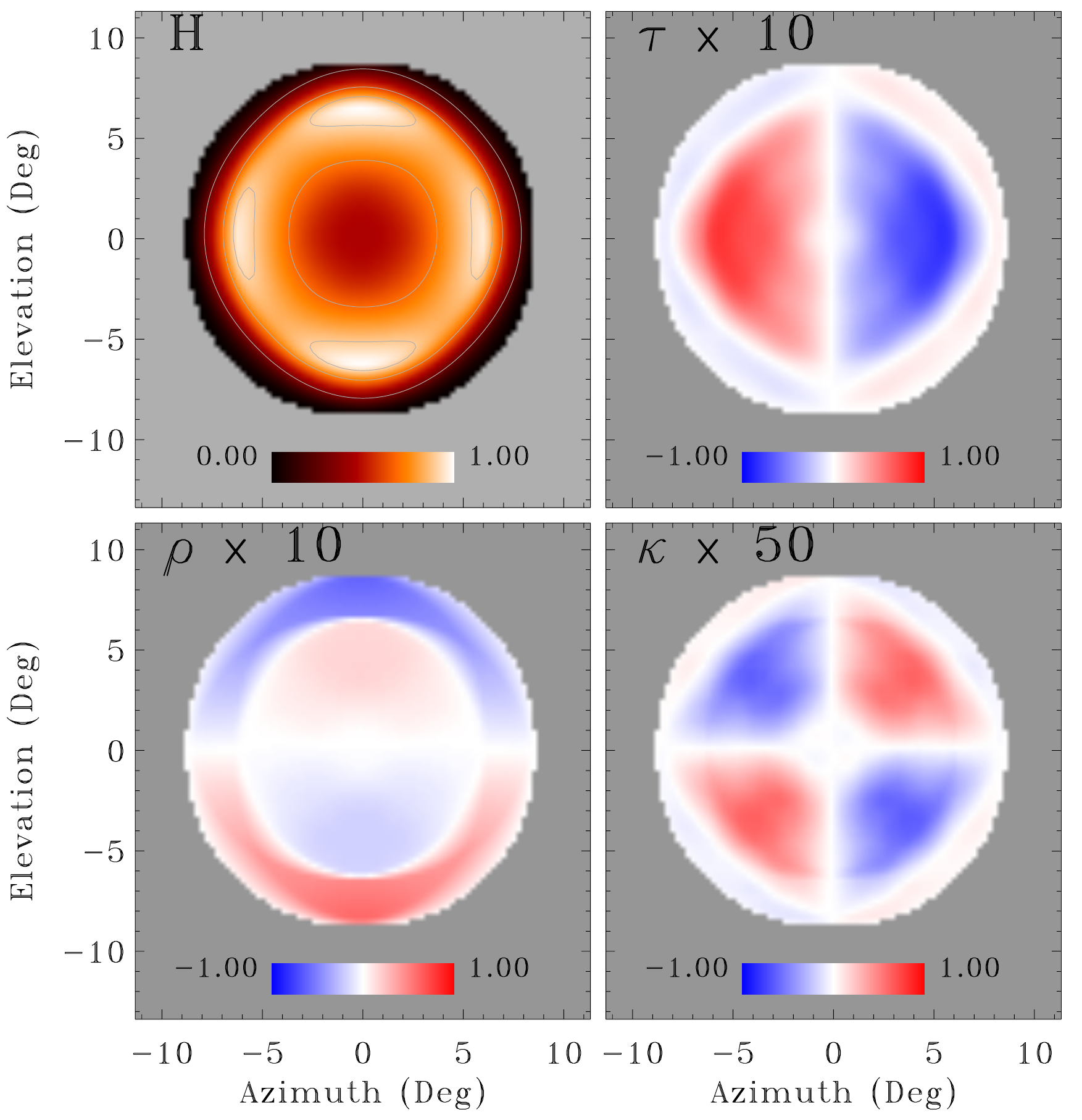}
\end{center}
\caption
{Common-mode and differential beam patterns
for the PIXIE feed horn concentrators.
The feed horn beam pattern
does not directly source $T \rightarrow B$ systematic errors,
but only modulates the effect from the differential fore-optics.
The off-axis design
creates dipolar modulation 
in the differential beam patterns $\rho$ and $\tau$,
while the square shape is reflected in the 
quadrupolar modulation for $\kappa$.
Contours for the common-mode response $H$
are shown at amplitude 0.3, 0.7, and 0.9.
\label{horn_beam_fig}
}
\end{figure}

Differences between the 
two polarizations $\hat{x}$ and $\hat{y}$ 
within a single concentrator can occur,
corresponding to the difference
between the E-plane and H-plane beam patterns
for a single-moded feed.
PIXIE's multi-moded operation reduces this effect,
which vanishes in the geometric optics limit.
We further reduce the effect by rotating the concentrator
so that the symmetry axes of the square aperture
lie at $\pm 45\deg$~
relative to the $\hat{x}$ and $\hat{y}$ polarization vectors
(Fig \ref{horn_cartoon}).
The resulting beams in $\hat{x}$ and $\hat{y}$
are equivalent linear combinations
of the E-plane and H-plane beam patterns,
so that
\begin{eqnarray}
H_{Lx} &\approx& H_{Ly}	\nonumber \\
H_{Rx} &\approx& H_{Ry} 
\label{horn_pol_eq}
\end{eqnarray}
with residuals resulting from 
small displacements in the rotation angle
\cite{kogut/fixsen:2018}.
Without loss of generality,
we may follow Eq. \ref{FAX_def}
to decompose the beam pattern from each horn
into a component common to all four detectors
plus a set of differential beam patterns,
\begin{eqnarray}
H        &=& \left( H_{Lx} + H_{Ly} + H_{Rx} + H_{Ry} \right) / 4  \nonumber \\
\rho     &=& \left( H_{Lx} - H_{Ly} + H_{Rx} - H_{Ry} \right) / 4	\nonumber \\
\tau     &=& \left( H_{Lx} + H_{Ly} - H_{Rx} - H_{Ry} \right) / 4	\nonumber \\
\kappa   &=& \left( H_{Lx} - H_{Ly} - H_{Rx} + H_{Ry} \right) / 4	
\label{H_def}
\end{eqnarray}

\noindent
so that
\begin{eqnarray}
H_{Lx} &=& H + \rho + \kappa + \tau \nonumber	\\
H_{Ly} &=& H - \rho + \kappa - \tau \nonumber	\\
H_{Rx} &=& H + \rho - \kappa - \tau \nonumber	\\
H_{Ry} &=& H - \rho - \kappa + \tau 
\label{horn_def}
\end{eqnarray}
where the horn parameters
are defined analogously to the fore-optics in Eq. \ref{F_def}.
Figure \ref{horn_beam_fig} 
shows the common-mode and differential beam patterns
from the concentrator horn.
Asymmetries from the off-axis orientation
appear at the few-percent level.
As with the fore-optics,
the differential beams are dominated
by an anti-symmetric (dipolar) component.

Appendix A shows the full response for each detector.
Retaining only terms to first order in the beam differences,
the signals at each detector become
\begin{eqnarray}
P_{Lx} = ~~{\boldsymbol Q} \, HF
	&+& 	{\boldsymbol Q} \, ( ~H \epsilon + F\rho + F\kappa + F\tau) \nonumber \\
	&+& 	{\boldsymbol I} \, H (\delta + \Delta)		\nonumber \\
%
%
P_{Ly} = -{\boldsymbol Q} \, HF
	&+& 	{\boldsymbol Q} \, ( ~H \epsilon + F\rho - F\kappa + F\tau ) \nonumber \\
	&-& 	{\boldsymbol I} \, H (\delta - \Delta)		\nonumber \\
%
%
P_{Rx} = ~~{\boldsymbol Q} \, HF
	&+& 	{\boldsymbol Q} \, ( -H \epsilon + F\rho - F\kappa + F\tau ) \nonumber \\
	&+& 	{\boldsymbol I} \, H (\delta - \Delta)		\nonumber \\
%
%
P_{Ry} = -{\boldsymbol Q} \, HF
	&+& 	{\boldsymbol Q} \, ( -H \epsilon + F\rho + F\kappa - F\tau ) \nonumber \\
	&-& 	{\boldsymbol I} \, H (\delta + \Delta) ~~.
\label{big_4_det_first}
\end{eqnarray}
Systematic errors from the concentrator beam pattern
appear in the second (polarization amplitude) term.
Although larger in amplitude 
than the spatial-polarization error ${\boldsymbol Q} \epsilon$,
these terms do not couple temperature to polarization
and so may be absorbed into the calibration.
The final term representing temperature--polarization mixing
is dominated at lowest order 
by the differential error from the fore-optics.

\section{Combined Detector Response}

PIXIE's four detectors
share different portions of the optical system
(left or right concentrator,
$\hat{x}$ or $\hat{y}$ polarization).
Linear combinations of the post-detection signals
can either eliminate or isolate specific systematic error signals,
providing additional safeguards against
temperature--polarization mixing.
For example, we may combine all 4 detectors 
to yield the sum signal
\begin{eqnarray}
\left[ P_{Lx} - P_{Ly} + P_{Rx} - P_{Ry} \right] / 4 &=&  			
	{\boldsymbol Q} \, HF			\nonumber \\
  &+&	{\boldsymbol Q} \, \epsilon \tau	\nonumber \\
  &+&	{\boldsymbol I} \, H \delta  		\nonumber \\
  &+&	{\boldsymbol I} \, \Delta \tau	~~,	
\label{4_det_coadd}
\end{eqnarray}
where we now retain terms to second order in the differential beam patterns.
As before, the first term is the true sky polarization,
convolved with the combined common-mode beam pattern
from the feed horn and fore-optics.
The second term affects only the amplitude
of the true sky polarization 
and may be absorbed into the calibration.
The final two terms
represent systematic errors
coupling temperature anisotropy to polarization.

We use Monte Carlo ray-trace simulations
to quantify the expected amplitude of these terms.
Table \ref{beam_summary_table}
summarizes the common-mode and differential beam patterns
for the PIXIE optical system.
The differential beam patterns
are small compared to the common-mode response.
We compare the weighted beam area of the differential beams
to the weighted area of the common-mode beam pattern,
\begin{equation}
f = \frac{ \int | \Delta(\theta,\phi)| d\Omega }
	 { \int |      F(\theta,\phi)| d\Omega } ~,
\label{fractional_area_eq}
\end{equation}
computed similarly for each of the 6 differential beam patterns.
The differential beams have fractional area
of a few percent for the concentrator,
and $10^{-2}$ to $10^{-5}$ for the more symmetric fore-optics.
The differential beams are dominated by a
dipolar modulation ($m=1$) which does not 
lead to temperature--polarization mixing.
The systematic error response to spin modulation at $m=2$
is typically of order $10^{-6}$ or smaller.

%
\begin{table}[t]
\caption{Spin Modulation of the Common-Mode and Differential Beam Patterns}
\label{beam_summary_table}
\begin{center}
\begin{tabular}{l c c c c}
\hline
\hline
  &  \multicolumn{4}{c}{Fore-Optics} \\
Parameter 		& $F$ 			    & $\Delta$			& $\delta$		  & $\epsilon$	\\
\hline
Peak Amplitude         &    1   		    &    $3 \times 10^{-2}$    &    $2 \times 10^{-4}$    &    $1 \times 10^{-4}$  \\
Relative Beam Area $f$ &    1    		    &    $2 \times 10^{-2}$    &    $8 \times 10^{-5}$    &    $4 \times 10^{-5}$  \\
Power (m=1)            &    $7 \times 10^{-5}$      &    $2 \times 10^{-5}$    &    $9 \times 10^{-11}$   &    $2 \times 10^{-10}$  \\
Power (m=2)            &    $3 \times 10^{-4}$      &    $1 \times 10^{-6}$    &    $4 \times 10^{-9}$    &    $7 \times 10^{-11}$  \\
\hline
\hline
  &  \multicolumn{4}{c}{Horn Concentrator} \\
Parameter 		& $H$ 			    & $\rho$		   	& $\tau$		& $\kappa$	\\
\hline
Peak Amplitude         &    1   		    &    $6 \times 10^{-2}$    &    $8 \times 10^{-2}$    &    $1 \times 10^{-2}$  \\
Relative Beam Area $f$ &    1    		    &    $3 \times 10^{-2}$    &    $5 \times 10^{-2}$    &    $7 \times 10^{-3}$  \\
Power (m=1)            &    $5 \times 10^{-8}$      &    $2 \times 10^{-4}$    &    $9 \times 10^{-4}$    &    $4 \times 10^{-8}$  \\
Power (m=2)            &    $6 \times 10^{-4}$      &    $8 \times 10^{-10}$   &    $4 \times 10^{-6}$    &    $3 \times 10^{-5}$  \\
\hline
\hline 
\end{tabular}
\end{center}
\end{table}

We may now quantify the systematic error terms
in the post-detection linear combination.
The third term
${\boldsymbol I} \, H \delta$
in Eq. \ref{4_det_coadd}
is similar to the temperature--polarization mixing
${\boldsymbol I} \, H \Delta$
from a single detector
(Eq. \ref{delta_diff}),
but reduced in amplitude by a factor of 200 
due to replacing the
A--B differential beam pattern $\Delta$
with the smaller
$\hat{x} - \hat{y}$ differential beam pattern $\delta$.
The lower response to $m=2$ modulation
from the $\delta$ differential beam
(compared to the $\Delta$ beam)
produces additional systematic error suppression.
The final term
${\boldsymbol I} \, \Delta \tau$
also represents temperature--polarization mixing,
but now appears at second order in small beam differences
and is reduced by a factor $20$ in amplitude
from the single-detector error.
The $m=2$ spin modulation of the $\tau$ differential beam
yields additional suppression.

We may also choose linear combinations of detectors to
cancel the polarized sky signal ${\boldsymbol Q} \, HF$,
thereby isolating
specific systematic error signals.
Such measurements of the systematic error signals
can be used
both to correct the sky measurements
and as confirmation of the expected effect
from beam pattern differences.
For example, the orthogonal combination of four detectors
becomes
\begin{eqnarray}
\left[ P_{Lx} - P_{Ly} - P_{Rx} + P_{Ry} \right] / 4 &=&  			
	{\boldsymbol Q} \, F \kappa		\nonumber \\
  &+&	{\boldsymbol Q} \, \epsilon \tau	\nonumber \\
  &+&	{\boldsymbol I} \, \Delta \rho		\nonumber \\
  &+&	{\boldsymbol I} \,  \delta \kappa ~~.	
\label{4_det_diff}
\end{eqnarray}
We may again use Table \ref{beam_summary_table} 
to estimate the amplitude of each term.
Unpolarized CMB signals ${\boldsymbol I}$
have amplitude of order 100 $\mu$K,
while the E-mode polarization ${\boldsymbol Q}$ is of order 1 $\mu$K.
Multiplying each CMB term
by the relative beam area of each beam pattern
yields an estimate of the relative amplitude of each term
(prior to spin modulation).
The difference signal
is dominated by the term
${\boldsymbol I} \, \Delta \rho$,
representing the convolution of the
unpolarized CMB anisotropy
with the double beam difference $\Delta \rho$. 
We may instead choose to compare 
signals from the two detectors
sharing a common concentrator.
A similar analysis shows that the detector-pair combination
$( P_{Lx} + P_{Ly} )/2$
is dominated by the term
${\boldsymbol I} \, H \Delta$
which isolates a single differential beam
for measurement and correction.
Similar linear combinations can isolate other terms.

\section{Tolerance}
Mirror symmetries within PIXIE's differential optics
suppress systematic errors coupling unpolarized structure
in the sky to a false polarized signal.
Positioning errors in the optical components
during assembly
can distort the beams from the ideal beam patterns.
We quantity the resulting degradation in optical performance
using 30 Monte Carlo realizations of the PIXIE optical system.
For each realization, 
we adjust the position of each optical element
allowing both translation and rigid-body rotation
about its nominal orientation
assuming assembly and machining tolerances of $\pm$0.05 mm
drawn from a random Gaussian distribution.
After adjusting all optical elements,
we follow the paths of $10^9$ rays through the
adjusted optical system
to define the distorted beam patterns.

\begin{figure}[t]
\begin{center}
\includegraphics[width=4.0in]{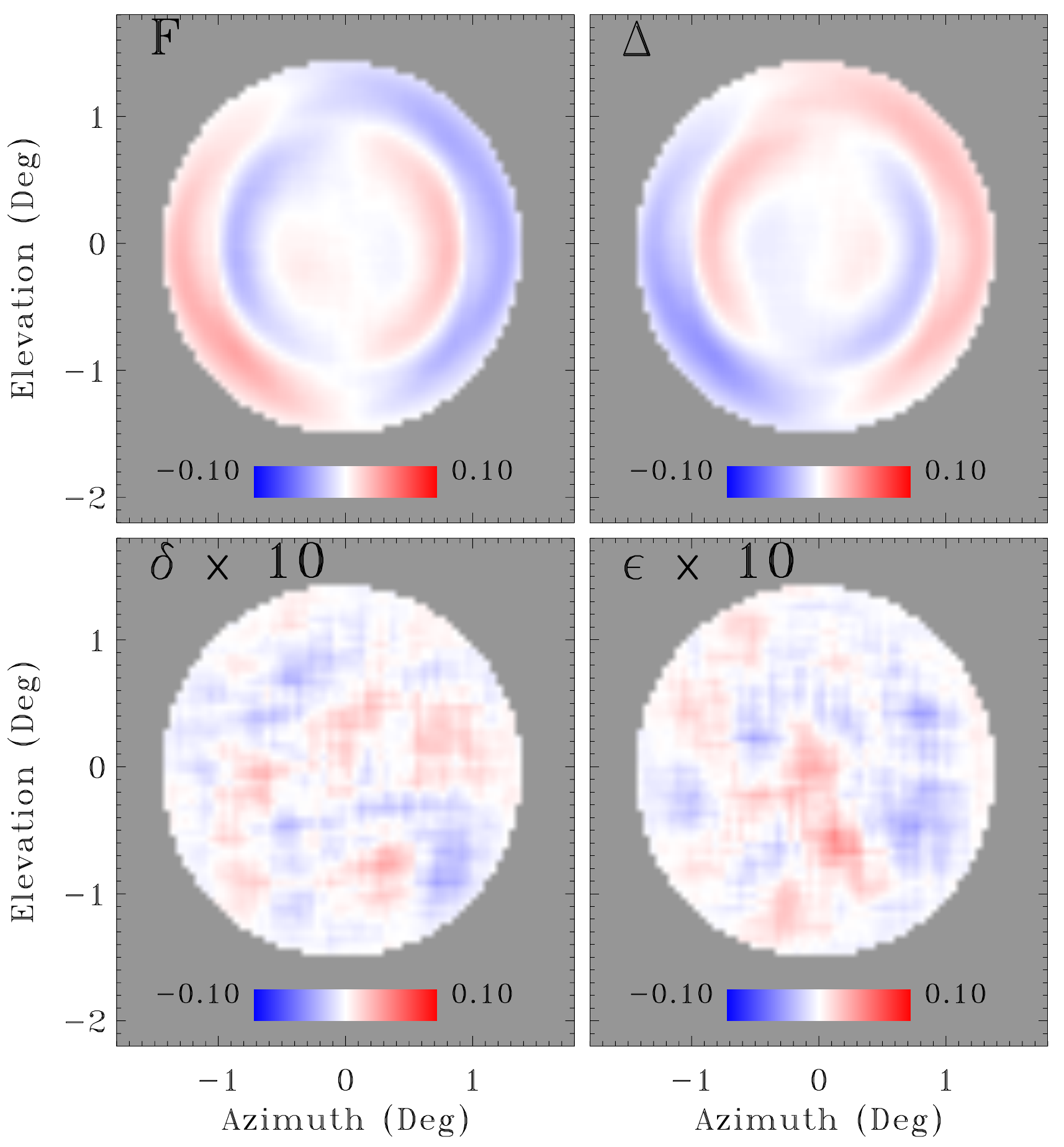}
\end{center}
\caption
{Differences between the nominal beam patterns
from Figure \ref{fore_optics_beams}
and the distorted beam patterns
after allowing for machining and assembly tolerances.
Beam patterns are shown
from a single Monte Carlo realization
in which the position and orientation of each optical element
are perturbed about the nominal configuration.
\label{fore_optics_diff_beams}
}
\end{figure}

The PIXIE optical system is robust to typical machining
and assembly tolerances.
The FTS left and right transfer mirror sets
and the mid-plane septum containing the polarizing grids
are each machined from a single block of aluminum.
The relative position and orientation of the mirrors or grids
within each set
have the $\pm$0.02 mm tolerance of computer-aided milling machines.
This minimizes relative displacement of these components during assembly
(although each set can still be displaced as a rigid body).
All optical elements
as well as the supporting structure
are fabricated from the same material (aluminum)
so that self-similar thermal contraction retains optical alignment.
Alignment of the primary, secondary, and folding flat mirrors
relative to the FTS
assumes somewhat looser $\pm$0.05 mm tolerance
typical of pinned construction.
Since these components are machined individually,
tolerancing errors should be uncorrelated.
Figure \ref{fore_optics_diff_beams} shows the difference
between the nominal beam patterns 
and the distorted patterns
for a single Monte Carlo realization of the 
distorted optical system.
The dominant effect is an angular displacement
of order 3$\amin$
between the $A$ and $B$ beams on the sky,
caused by displacements of the primary and secondary mirrors.
This in turn creates
an anti-symmetric (dipolar) pattern
in both the mean beam ($F$)
and the A--B spatial asymmetry ($\Delta$).
Angular displacement of the beam centroid
couples to spin moment $m=1$ 
and does not induce additional temperature--polarization mixing.

\begin{figure}[t]
\begin{center}
\begin{tabular}{c}
\includegraphics[height=2.6in]{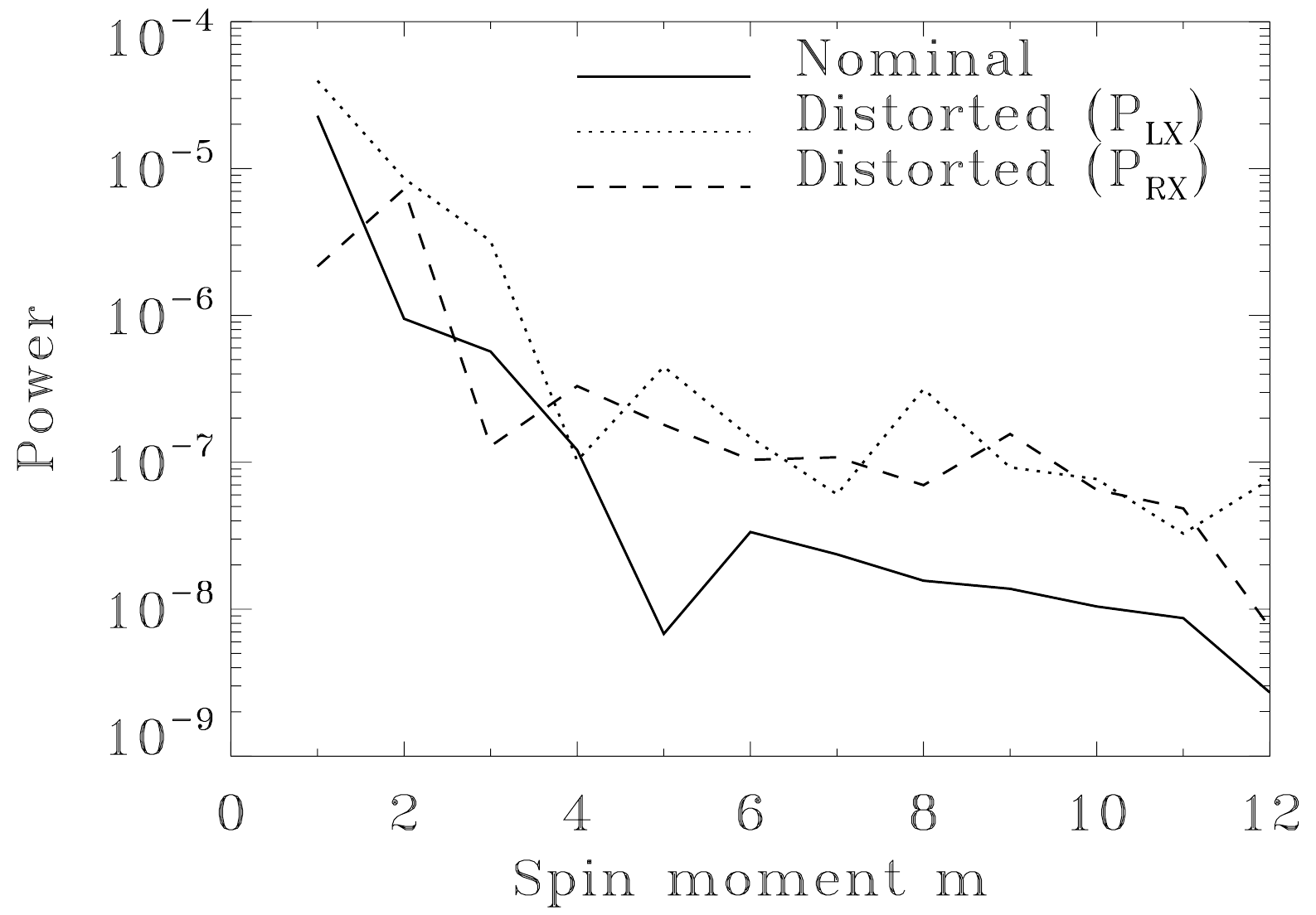}
\end{tabular}
\end{center}
\caption
{Effects of machining and assembly tolerances on the differential beam pattern
$\Delta(\theta, \phi$).
We compare the differential beam pattern
for the nominal optical configuration
to a Monte Carlo realization
with all optical elements perturbed
from their nominal positions.
The distorted patterns are shown 
as a function of spin moment $m$
for the detectors sensitive to $\hat{x}$ sky polarization
in both the left-side and right-side concentrators.
The distorted optical system
still shows suppression of order $10^{-5}$
for temperature-polarization coupling at $m=2$.
\label{distorted_beams_vs_m}
}
\end{figure}

Figure \ref{distorted_beams_vs_m}
compares the spin dependence of the nominal beam patterns
to the distorted beams
generated from a single Monte Carlo realization 
of the full optical system.
It is similar to the ideal beam patterns
shown in Figure \ref{fore_optics_vs_m},
except that the position and pointing
of each optical element
has been perturbed by an amount
randomly chosen from a Gaussian normal distribution
of width 0.05 mm.
We now also include the illumination of the (perturbed) fore-optics
by the (perturbed) concentrator.
For clarity, we compare the spin decomposition
for the nominal and distorted configurations
for a single choice of differential beam.
Temperature-polarization coupling 
for a single detector is dominated by
the A--B differential beam $\Delta(\theta,\phi)$
(Eq. \ref{big_4_det_first}).
PIXIE has 4 detectors;
we show the distorted beam decomposition
for detectors observing the same ($\hat{x}$) sky polarization
from either the left or right concentrator.
Compared to the ideal system,
the distorted optical system has a larger response
to systematic error coupling at $m=2$,
but the response is still suppressed by
a factor of $10^5$ compared to the true sky polarization.

\section{Discussion}

Systematic errors coupling unpolarized anisotropy 
to a false polarized signal
are a common concern to CMB polarimeters.  
PIXIE's optical design provides several layers 
of mitigation
compared to instruments 
imaging the CMB across a large (kilo-pixel) focal plane.
Missions employing kilo-pixel arrays
across large fields of view
must account for the systematic degradation
in beam shape
from coma and shear
for detectors farther from the center of the focal plane.
All four PIXIE detectors,
in contrast,
lie at the center of the focal plane,
allowing eam deformation to be minimized
(Fig \ref{fore_optics_beams}).

More importantly,
PIXIE's four-port optical system provides 
three distinct levels of differential measurement.
The Fourier transform spectrometer
produces a signal that depends
on the difference 
between two nearly-identical beams on the sky.
This differential measurement is performed optically,
prior to detection,
and is independent of detector calibration.
We use ray-trace simulations to evaluate 
the differential beam patterns
after removing the common-mode response.
The differential beams
can be described in terms of
the spatial asymmetry between the A- and B-sides of the instrument,
the polarization asymmetry between the $\hat{x}$ and $\hat{y}$ response,
plus a cross term for the mixed
spatial-polarization difference.
All of the differential beams are small 
compared to the common-mode response.
The largest effect is the spatial (A--B) asymmetry,
which has only 1.5\% of the common-mode response.
The other differential beams have response below 0.01\%.

PIXIE's symmetric design further reduces 
systematic error response from the differential beam cancellation.
The FTS interferes the $\hat{x}$ polarization from the A-side beam
with the $\hat{y}$ polarization from the B-side beam
(Fig \ref{pol_symmetry_cartoon}).
The optical system is symmetric about the mid-plane
between the two sides,
which forces the 
$\hat{x}$ polarization from one beam
to be the mirror reflection of the
$\hat{x}$ polarization from the other beam.
The A--B mirror reflection
combines with the
A--B beam subtraction
to produce an anti-symmetric (dipole) response
in the differential beam patterns.
The anti-symmetric part of the differential beam pattern
does not contribute to the systematic error
from temperature-polarization coupling.
Each detector samples a single polarization state;
the entire instrument spins about the boresight
to allow full sampling of the sky polarization.
True polarized signals appear at twice the spin frequency,
while anti-symmetric signals
can only appear at odd harmonics of the spin.
Systematic errors from temperature--polarization coupling
thus depend only on the
$m=2$ component of the differential beam patterns,
which are dominated by a dipole ($m=1$) response.
Ray-trace models of the PIXIE beams
show that the response at $m=2$
is reduced by an additional factor of $10^6$ or more.
In principle, the optical system could further be optimized
to suppress the $m=2$ differential beam response,
moving power to other $m$ values
that do not participate in temperature--polarization mixing.
This has not yet been done
but is planned for future development.

Finally, we may follow the common practice for CMB measurements
and combine the post-detection signals from individual detectors.
The four detectors are mounted in identical concentrators
and view the same sky direction
through the same optical path.
Combining all four detectors
cancels the leading effects
from differential beams in the single-detector signal,
reducing the systematic error response 
by a factor of 1000 or more
compared to the individual detectors.
Conversely, 
orthogonal linear combinations of
2 or 4 detectors
can cancel the polarized sky signal
to isolate, identify, and model
specific systematic effects from
the individual differential beam patterns.

Systematic error suppression in the differential PIXIE optics
is robust against typical machining and assembly tolerances.
We combine ray-trace optical simulations
with 
Monte Carlo realizations of distorted PIXIE optics
to evaluate 
both the individual beam patterns
and the resulting systematic error response.
Each Monte Carlo realization of then PIXIE optics
perturbs each optical element
(mirrors, folding flats, polarizing grids, etc)
in both position and orientation
by an amount drawn from a Gaussian distribution
whose width is set by typical machining/assembly tolerances
of 0.05 mm.
The dominant effect of such tolerance errors
is an angular displacement of the A-side beams
relative to the B-side beams.
The two beams are normally co-pointed on the sky;
after accounting for tolerances
the beams are typically mis-aligned by 3$\amin$.
This is small compared to the 2.6\deg 
~width of the common-mode beams;
the resulting dipolar beam asymmetries
predominantly effect the $m=1$ spin moment
and do not couple efficiently to polarization.
The distorted optical system
still provides suppression of the $m=2$
temperature--polarization systematic error
by factor
of order $10^5$.

\appendix

\section{Full Single-Detector Systematic Error}
Expanding Eqs. \ref{4_det_with_delta} and \ref{horn_def}
yields individual detector signals
\begin{eqnarray}
P_{Lx} = ~~{\boldsymbol Q} \, HF
	& & ~~~~~~~~
	~~+~~ 	{\boldsymbol Q} \, H \epsilon
	 ~+~~ 	{\boldsymbol I} \, H \delta
	 ~+~~ 	{\boldsymbol I} \, H \Delta		\nonumber \\
	&+& 	{\boldsymbol Q} \, F \rho
	~~+~~ 	{\boldsymbol Q} \,  \rho \epsilon
	~~+~~ 	{\boldsymbol I} \, \rho \delta
	~~+~~ 	{\boldsymbol I} \, \rho \Delta		\nonumber \\
	&+& 	{\boldsymbol Q} \, F \kappa
	~~+~~ 	{\boldsymbol Q} \,  \kappa \epsilon
	~~+~~ 	{\boldsymbol I} \, \kappa \delta
	~~+~~ 	{\boldsymbol I} \, \kappa \Delta	\nonumber \\
	&+& 	{\boldsymbol Q} \, F \tau
	~~+~~ 	{\boldsymbol Q} \,  \tau \epsilon
	~~+~~ 	{\boldsymbol I} \, \tau \delta
	~~+~~ 	{\boldsymbol I} \, \tau \Delta	\\
	& &	\nonumber \\
P_{Ly} = -{\boldsymbol Q} \, HF
	& & ~~~~~~~~
	~~+~~ 	{\boldsymbol Q} \, H \epsilon
	 ~-~~ 	{\boldsymbol I} \, H \delta
	 ~+~~ 	{\boldsymbol I} \, H \Delta		\nonumber \\
	&+& 	{\boldsymbol Q} \, F \rho
	~~-~~ 	{\boldsymbol Q} \,  \rho \epsilon
	~~+~~ 	{\boldsymbol I} \, \rho \delta
	~~-~~ 	{\boldsymbol I} \, \rho \Delta		\nonumber \\
	&-& 	{\boldsymbol Q} \, F \kappa
	~~+~~ 	{\boldsymbol Q} \,  \kappa \epsilon
	~~-~~ 	{\boldsymbol I} \, \kappa \delta
	~~+~~ 	{\boldsymbol I} \, \kappa \Delta	\nonumber \\
	&+& 	{\boldsymbol Q} \, F \tau
	~~-~~ 	{\boldsymbol Q} \,  \tau \epsilon
	~~+~~ 	{\boldsymbol I} \, \tau \delta
	~~-~~ 	{\boldsymbol I} \, \tau \Delta	\\
	& &	\nonumber \\
P_{Rx} = ~~{\boldsymbol Q} \, HF
	& & ~~~~~~~
	~~-~~ 	{\boldsymbol Q} \, H \epsilon
	 ~+~~ 	{\boldsymbol I} \, H \delta
	 ~-~~ 	{\boldsymbol I} \, H \Delta		\nonumber \\
	&+& 	{\boldsymbol Q} \, F \rho
	~~-~~ 	{\boldsymbol Q} \,  \rho \epsilon
	~~+~~ 	{\boldsymbol I} \, \rho \delta
	~~-~~ 	{\boldsymbol I} \, \rho \Delta		\nonumber \\
	&-& 	{\boldsymbol Q} \, F \kappa
	~~+~~ 	{\boldsymbol Q} \,  \kappa \epsilon
	~~-~~ 	{\boldsymbol I} \, \kappa \delta
	~~+~~ 	{\boldsymbol I} \, \kappa \Delta	\nonumber \\
	&-& 	{\boldsymbol Q} \, F \tau
	~~+~~ 	{\boldsymbol Q} \,  \tau \epsilon
	~~-~~ 	{\boldsymbol I} \, \tau \delta
	~~+~~ 	{\boldsymbol I} \, \tau \Delta	\\
	& &	\nonumber \\
P_{Ry} = -{\boldsymbol Q} \, HF
	& & ~~~~~~~
	~~-~~ 	{\boldsymbol Q} \, H \epsilon
	 ~-~~ 	{\boldsymbol I} \, H \delta
	 ~-~~ 	{\boldsymbol I} \, H \Delta		\nonumber \\
	&+& 	{\boldsymbol Q} \, F \rho
	~~+~~ 	{\boldsymbol Q} \,  \rho \epsilon
	~~+~~ 	{\boldsymbol I} \, \rho \delta
	~~+~~ 	{\boldsymbol I} \, \rho \Delta		\nonumber \\
	&+& 	{\boldsymbol Q} \, F \kappa
	~~+~~ 	{\boldsymbol Q} \,  \kappa \epsilon
	~~+~~ 	{\boldsymbol I} \, \kappa \delta
	~~+~~ 	{\boldsymbol I} \, \kappa \Delta	\nonumber \\
	&-& 	{\boldsymbol Q} \, F \tau
	~~-~~ 	{\boldsymbol Q} \,  \tau \epsilon
	~~-~~ 	{\boldsymbol I} \, \tau \delta
	~~-~~ 	{\boldsymbol I} \, \tau \Delta	
\label{big_4_det_2nd}
\end{eqnarray}
where now we retain all terms to second order.

%


\vspace{2ex}
\noindent
\textbf{Alan Kogut} is an astrophysicist at NASA's Goddard Space Flight Center.
He received his A.B. from Princeton University
and his PhD from the University of California at Berkeley.
His research focuses on observations of the
frequency spectrum and linear polarization
of the cosmic microwave background
and
diffuse astrophysical foregrounds
at millimeter and sub-mm wavelengths.

\vspace{2ex}
\noindent
\textbf{Dale Fixsen} is an astrophysicist at the University of Maryland. 
He received his B.S. from Pacific Lutheran University, Tacoma Washington
and his PhD from Princeton University. 
His research focuses on the spectrum, temperature, and polarization 
of the cosmic microwave background 
and the radio and infrared cosmic backgrounds.

\end{spacing}
\end{document}